\documentclass[12pt]{iopart}

\usepackage{iopams}

\expandafter\let\csname equation*\endcsname\relax
\expandafter\let\csname endequation*\endcsname\relax

\usepackage{amsmath}
\usepackage{graphicx}
\usepackage{hyperref}
\usepackage{subcaption}
\usepackage{color}

\def\bra#1{\mathinner{\langle{#1}|}}
\def\ket#1{\mathinner{|{#1}\rangle}}

\begin{document}

\title[There is no general connection between the QSL and non-Markovianity]{There is no general connection between the quantum speed limit and non-Markovianity}

\author{J Teittinen$^{1,2}$, H Lyyra$^{1,2}$ and S Maniscalco$^{1,2,3}$}
\address{$^1$ Turku Centre for Quantum Physics, Department of Physics and Astronomy, University of Turku, FIN-20014, Turun Yliopisto, Finland}
\address{$^2$ QTF Centre of Excellence, Department of Physics and Astronomy, University of Turku, FI-20014 Turun Yliopisto, Finland}
\address{$^3$ QTF Centre of Excellence, Department of Applied Physics, Aalto University, FIN-00076 Aalto, Finland}
\ead{jostei@utu.fi}

\begin{abstract}
\noindent 
The quantum speed limit sets a bound on the minimum time required for a quantum system to evolve between two  states. For open quantum systems this quantity depends on the dynamical map describing the time evolution in presence of the environment, on the evolution time $\tau$, and on the initial state of the system. We consider a general single qubit open dynamics and show that there is no simple relationship between memory effects and the tightness of the quantum speed limit bound. We prove that only for specific classes of dynamical evolutions and initial states, there exists a link between non-Markovianity and the quantum speed limit. Our results shed light on the connection between information back-flow between system and environment and the speed of quantum evolution.
\end{abstract}

\maketitle

\section{Introduction}

The idea of the possible existence of a fundamental limit, dictated by the principles of quantum mechanics, for the speed of evolution of quantum states was firstly discussed in Ref.~\cite{mandelstam1945}. In that paper, Mandelstam and Tamm derived a quantum speed limit (QSL) from the Heisenberg energy-time uncertainty relation. Specifically, they showed that the evolution time is bounded by the variance of energy as $\tau_{MT} \geq h/4\Delta E$. More recently, Margolus and Levitin studied the QSL in connection to the maximal rate of computation possible for a computer \cite{margolus1998}. In this case, the QSL was calculated as the minimum time for a quantum system to evolve from a pure initial state to some orthogonal pure state, using a one dimensional harmonic oscillator as an example. The authors showed that the minimum time is related to the total energy of the system as $\tau_{ML} \geq h/4E$. These two bounds are unordered, and therefore in the literature the QSL is  defined as the maximum between these two quantities. 

The results of Refs.~\cite{mandelstam1945,margolus1998} were extended to include cases where the evolved state is not orthogonal to the initial state in Ref.~\cite{giovannetti2003}. Moreover, in addition to the previous definitions valid for closed quantum systems, several authors proposed different generalizations to open quantum systems applicable for both Markovian and non-Markovian dynamics \cite{deffner2013,huelga2013,matos2013,pires2016,deffner2017b}. Nowadays, QSLs are investigated in connection to a number of topics, from quantum metrology to quantum computation, from quantum control to quantum thermodynamics, as reviewed, e.g., in Ref.~\cite{deffner2017}. Contrarily to what was initially believed, speed limits are not an exclusive property of quantum systems, namely they do not arise uniquely because of quantum features. Indeed, they can be derived also for classical systems, without assuming any quantum properties, such as commutation relations, as shown recently in \cite{okuyama2018} and \cite{shanahan2018}.

In this paper we focus on the geometric formulation of the quantum speed limit given in Ref.~\cite{deffner2013}. We are specifically interested in clarifying the connection between the QSL bound and the presence or absence of memory effects, described in terms of information backflow \cite{breuer2009}. Following Ref.~\cite{deffner2013}, this aspect has been further investigated, elaborating on the claim that the QSL is smaller when the dynamics is non-Markovian, potentially speeding up the evolution \cite{xu2014,wu2015}. These authors showed analytically that, for a specific model of open quantum system dynamics, the ratio between the QSL and the actual evolution time, $\tau_{QSL}/\tau$, is 1 when the system is Markovian, and is smaller than 1 when it is non-Markovian. Their result suggests that in the Markovian case the dynamics saturates the bound, giving the most efficient evolution, whereas in the non-Markovian case the actual limit can still be lower than the evolution time. The explicitly derived dependency between QSL and non-Markovianity has proven useful in several applications \cite{deffner2017,wu2015,zhang2015,liu2016,zhang2016,cai2017,xu2018,xu2018b,liu2015,hou2016,song2016,mo2017,wu2016,wang2016,song2017}.

Our main goal is to tackle the question of the connection between non-Markovianity and the quantum speed limit not starting from a specific model but in full generality, looking in detail at the role played by the dynamical map, the evolution time $\tau$, and the initial state, in the achievement of the QSL bound. 
We show that, for the most general cases, there is no simple connection between the Markovian to non-Markovian crossover and the QSL. Under certain more restrictive assumptions, however, we can characterize families of one-qubit dynamical maps for which the QSL speed-up coincides with the onset of non-Markovianity, as indicated by the Breuer-Laine-Piilo (BLP) non-Markovianity measure \cite{breuer2009}. For these families we derive analytical formulas for the QSL as a function of the BLP measure. Our results also show that, for a given open quantum system model, both the evolution time $\tau$ and the initial state play a key role and cannot be overlooked when making claims on the QSL. As an example, we generalise results in \cite{deffner2013} to a broader set of pure initial states, and show that the QSL bound is saturated only for very few initial states even in the fully Markovian case. 

The paper is structured as follows. In Section \ref{sec:general_QSL} we briefly present the formalism of open quantum systems and recall the common mathematical definitions of QSL. In Section \ref{sec:jaynes_cummings} we present the Jaynes-Cummings model used in \cite{deffner2013} and discuss briefly their results concerning non-Markovianity and quantum speed-up. In Section \ref{sec:smalltau} we study how the actual evolution time affects the QSL for the same Jaynes-Cummings system. In Section \ref{sec:blp_and_qsl}, we calculate the general conditions for the QSL optimal dynamics, and study the connection between BLP non-Markovianity and QSL. In Section \ref{sec:markovian_mes} we study the initial state dependence of the QSL for the Markovian dynamics arising from Pauli and phase-covariant master equations. In Section \ref{sec:timedependent} we study the effects of Markovian to non-Markovian transition to QSL using a specific phase-covariant system as an example. In Section \ref{sec:conclusions} we summarize the results and discuss their implications.

\section{Quantum speed limit, non-Markovianity, and open quantum systems}\label{sec:general_QSL}

An open quantum system is a system ($S$) interacting with another system, the environment ($E$). Commonly the dynamics of $E$ is not interesting, and one concentrates only on how $S$ changes in time. In our case the system of interest is a single qubit. According to the theory of open quantum systems, the reduced dynamics of the qubit is given by $\rho_S(t) = \Phi_t(\rho_S(0)) = \text{tr}_E[U_{SE}^\dagger(t) \rho_{S}(0)\otimes \rho_{E}(0) U_{SE}^{ }(t)]$, where $\rho_S(t)$ is the reduced state of the system,  $\Phi_t$ the dynamical map, $\rho_{S}(0)\otimes \rho_{E}(0)$ the initial combined system-environment state, $U_{SE}(t)$ the unitary time evolution of the combined system, and $\text{tr}_E[~\cdot~]$ the partial trace over the environment.

We call a map $k$-positive, if the composite map $\Phi_t \otimes \mathbb{I}_k$, where $\mathbb{I}_k$ is the identity map of a $k$-dimensional ancillary Hilbert space, is positive for all $t \geq 0$. If a map is 1-positive, that is $k=1$, we call it a positive (P) map. If the map is $k$-positive for all $k\geq 0$, then we call the map completely positive (CP). Furthermore, a map is called CP-divisible (P-divisible), if, for any two time instants $s$ and $t$, with $s \geq t \geq 0$, the map can be written as
\begin{equation}
\Phi_s = V_{s,t} \circ \Phi_t \,,
\end{equation}
where the propagator $V_{s,t}$ is completely positive (positive).

The explicit models of dynamics in this paper are generated by a time-local master equation:
\begin{equation}\label{eq:lindblad_me}
\frac{d \rho_S(t)}{dt} = L_t(\rho_S(t)) =\frac{i}{\hbar}[\rho_S(t),H(t)] + \sum_i \gamma_i(t) \left( A_i^{ } \rho_s(t) A_i^\dagger - \frac{1}{2}\left\lbrace A_i^\dagger A_i^{ }, \rho_S(t) \right\rbrace \right)\,,
\end{equation}
where $H$ is the system Hamiltonian, $\gamma_i(t)$ the time-dependent decay rates, and $A_i$ the Lindblad operators. The solution for the master equation gives the time evolution of the state in the form of a dynamical map, $\Phi_t(\rho(0)) = \rho(t)$. The GKSL theorem implies, that for non-negative decay rates, that is $\gamma_i(t) \geq 0$, the resulting map is always completely positive and trace preserving (CPTP) \cite{gorini1976,lindblad1976,rivas2012}. CPTP is an important property, since it guarantees the physicality of the dynamical map.

The example dynamics considered in this paper arise from two very general families of master equations, namely the phase-covariant master equation \cite{lankinen2016,smirne2016,haase2018,teittinen2018}:
\begin{equation}\label{eq:pahsecovariant_me}
\begin{split}
L_t\rho_t &=~ i \omega(t) [\rho_t, \sigma_3] + \frac{\gamma_1(t)}{2} \left(\sigma_+ \rho_t \sigma_- - \frac{1}{2} \left\{ \sigma_- \sigma_+,\rho_t \right\} \right) \\
&~~+ \frac{\gamma_2(t)}{2} \left(\sigma_- \rho_t \sigma_+ - \frac{1}{2} \left\{\sigma_+ \sigma_-,\rho_t \right\} \right) + \frac{\gamma_3(t)}{2} \left(\sigma_3 \rho_t \sigma_3 - \rho_t\right) \,,
\end{split}
\end{equation} 
and the Pauli master equation \cite{andersson2007,Chruscinski2013}:
\begin{equation}\label{eq:pauli_me}
L_t \rho_t = \sum_{i=1}^3 \gamma_i(t) (\sigma_i \rho_t \sigma_i - \rho_t) \,,
\end{equation}
where $\sigma_1, \sigma_2$ and $\sigma_3$ are the Pauli $x$, $y$, and $z$ matrices respectively and $\sigma_{\pm} = \tfrac{1}{2}(\sigma_1 \pm i \sigma_2)$.

To study the effects of non-Markovianity, we employ the well-known BLP measure \cite{breuer2009}, defined as
\begin{equation}
\mathcal{N}(\Phi^{}_\tau) = \max_{\rho_1(0),\rho_2(0)} \int_{\sigma > 0} \sigma (\rho_{1,2},\Phi_{t}) dt \,,
\end{equation}
with $\sigma(\rho_{1,2},\Phi_t) = \tfrac{d}{dt}D(\Phi_t(\rho_1(0)),\Phi_t(\rho_2(0)))$, where $D(\rho_1(t),\rho_2(t)) = \tfrac{1}{2} \text{tr}\vert \rho_1(t)-\rho_2(t) \vert$ is the trace distance between $\rho_1(t)$ and $\rho_2(t)$ and the maximum is taken over all possible initial states, and the integral is calculated over $t \in (0,\tau)$. In this case, for $\mathcal{N} > 0$, the non-Markovianity is related to the amount of information flowing black to the system, quantified by the increase in distinguishability between the states. In terms of the dynamical map, this implies violation of P-divisibility \cite{breuer2009}.

The generalized quantum speed limit is defined as \cite{deffner2013}
\begin{equation}\label{eq:qsl_with_max}
\tau_{QSL} = \max \Big\lbrace \frac{1}{\Lambda_\tau^{op}} , \frac{1}{\Lambda_\tau^{tr}} , \frac{1}{\Lambda_\tau^{hs}} \Big\rbrace \sin^2 ( \mathcal{L}(\rho_0 , \rho_\tau) ),
\end{equation}
with $\mathcal{L}(\rho_0,\rho_\tau)$ the Bures angle between the pure initial state $\rho_0$ and the evolved state $\rho_\tau$, defined as
\begin{equation}
\mathcal{L}(\rho_0,\rho_\tau) := \arccos [\sqrt{F(\rho_0, \rho_\tau)}] \,,
\end{equation}
where $F(\rho_0, \rho_\tau) = (\text{tr}[\sqrt{\sqrt{\rho_0} \rho_\tau \sqrt{\rho_0}}])^2$ is the fidelity between the two states, which for pure initial state $\rho_0 = \ket{\psi_0}\bra{\psi_0}$ simplifies to
\begin{equation}
\mathcal{L}(\rho_0,\rho_\tau) = \arccos ( \sqrt{\langle \psi_0 \vert \rho_\tau \vert \psi_0 \rangle} )\,.
\end{equation}\label{eq:pure_buresangle}
We have denoted
\begin{equation}
\Lambda_\tau^{xx} = \frac{1}{\tau} \int_0^\tau || L_t(\rho_t) ||_{xx} dt \,,
\end{equation}
where $xx$ is either $op$, $tr$ or $HS$ for operator, trace, and Hilbert-Schmidt norm respectively. It can easily be shown, using the definitions 
\begin{align}
|| L_t \rho_t ||_{op} = \max_i \lbrace s_i \rbrace \,,~~~~
|| L_t \rho_t ||_{tr} = \sum_i s_i \,,~~~~
|| L_t \rho_t ||_{HS} = \sqrt{\sum_i s_i^2} \,,
\end{align}
where $s_i$ are the singular values of $L_t\rho_t$, that the operator norm always maximizes Eq.~\eqref{eq:qsl_with_max}, and thus the quantum speed limit can be written as
\begin{equation}
\tau_{QSL} = \frac{\sin^2 ( \mathcal{L}(\rho_0 , \rho_\tau) )}{\Lambda_\tau^{op}} \,.
\end{equation}

\section{Damped Jaynes-Cummings model}\label{sec:jaynes_cummings}

For the sake of concreteness we begin our investigation with a simple paradigmatic open quantum system model, extensively studied in the literature, which is a special case of the phase-covariant master equation given in \eqref{eq:pahsecovariant_me}. This  allows us to recall the results previously obtained in Ref. \cite{deffner2013}. We will then proceed to generalize these results along different lines, using this model for benchmarking.

The model considered is the resonant damped Jaynes-Cummings (JC) model, which can be obtained through an exact microscopic derivation from a total Hamiltonian describing a two-level system interacting with an infinite bosonic environment, e.g., the quantized field inside a leaky cavity. The dynamics of the two-level system is given by the master equation \cite{breuer1999}
\begin{equation}
L^{JC}_t(\rho_t) 
= \gamma(t)\left( \sigma_- \rho_t \sigma_+ - \frac{1}{2} \left\{ \sigma_+ \sigma_-, \rho_t \right\} \right)\,,
\end{equation}
with
\begin{equation}
\gamma(t) = \frac{2 \gamma_0 \lambda \sinh(dt/2)}{d \cosh(dt/2) + \lambda \sinh(dt/2)}\,,
\end{equation}
where $d = \sqrt{\lambda^2 - 2 \gamma_0 \lambda}$, $\lambda$ is the spectral width of the reservoir (hereafter assumed to be Lorentzian), and $\gamma_0$ is the coupling strength between the qubit and the cavity field. 
The solution to this system can be given in the following form
\begin{equation}\label{eq:JCmap}
\Phi^{JC}_t (\rho_0) = \rho_t = \left( \begin{array}{cc}
\rho_{11}|b_t|^2 & \rho_{10}b_t \\
\rho_{01}b_t^* & 1 - \rho_{11}|b_t|^2
\end{array} \right)\,,
\end{equation}
where $\rho_{11}$ corresponds to the excited state, and
\begin{equation}
b_t = e^{-\lambda t/2}\left( \cosh (dt/2) + \frac{\lambda}{d} \sinh(dt/2) \right) \,.
\end{equation}

In \cite{xu2014} it was numerically shown that for the map of Eq.~\eqref{eq:JCmap} the eigenstates $\vert 0 \rangle \langle 0 \vert$ and $\vert 1 \rangle \langle 1 \vert $ of $\sigma_3$,  are the optimal pair of states for the BLP measure. The trace distance for this pair is $D(\Phi^{}_t(\vert 0 \rangle \langle 0 \vert),\Phi^{}_t(\vert 1 \rangle \langle 1 \vert)) = |b_t|^2$, and so the BLP measure takes the form
\begin{equation}
\mathcal{N}(\Phi^{JC}_\tau) = \int_{\partial_t |b_t|^2 > 0} \partial_t |b_t|^2 dt \,.
\end{equation}
Following the calculations of Ref.~\cite{xu2014}, we can isolate the positive part of the integral by writing the integrand as $\partial |b_t|^2 = \tfrac{1}{2}(|\partial_t |b_t|^2 | + \partial_t |b_t|^2)$. Now the BLP measure can be written as an integral of the interval $[0,\tau]$ as
\begin{align}\label{eq:blpmeasure}
\mathcal{N}(\Phi^{JC}_\tau) &=   \frac{1}{2}\int_0^\tau \vert \partial_t |b_t|^2 \vert ~dt + \frac{1}{2}\big( |b_t|^2 - 1 \big) \,.
\end{align}

\noindent Choosing the initial state as $\vert 1 \rangle \langle 1 \vert$, the operator norm for the JC model becomes
\begin{equation}\label{eq:opmeasure}
|| L_t \rho_t ||_{op} = 
|\partial_t |b_t|^2| \,.
\end{equation}
Using Eqs. \eqref{eq:blpmeasure} and \eqref{eq:opmeasure}, and the identity $\sin^2 ( \arccos (f(t))) = 1 - f(t)^2$, we can write the QSL time as:
\begin{align}\label{eq:BLP_form}
\tau_{QSL} 
&= \frac{\tau}{\frac{2\mathcal{N}(\Phi^{JC}_\tau)}{1 - |b_\tau|^2} + 1} \,.
\end{align}

This equation suggests that the saturation of the QSL bound is strictly a feature of Markovian dynamics, since any dynamics with $\mathcal{N} (\Phi_\tau^{JC}) >0$ results in lower than optimal QSL. However, as we will show in the following, this consideration is valid only for dynamics described by Eq.~\eqref{eq:JCmap} 
and it cannot be used to describe QSL for other initial states. In what follows, we will generalize Eq.\eqref{eq:BLP_form}, firstly derived in Ref.~\cite{xu2014}, to a larger class of qubit dynamics and show that it does not hold in general. We also consider the QSL optimality of pure initial states which do not maximize the BLP measure.

\section{Evolution time dependence of $\tau_{QSL}/ \tau$}\label{sec:smalltau}

In this short section we show the dependence of the $\tau_{QSL}$ on the choice of the evolution time. More specifically we will see that $\tau_{QSL}$ is not monotonically dependent on  $\tau$ in the non-Markovian region.



\begin{figure}[!ht]\centering
\includegraphics[width=0.5\textwidth]{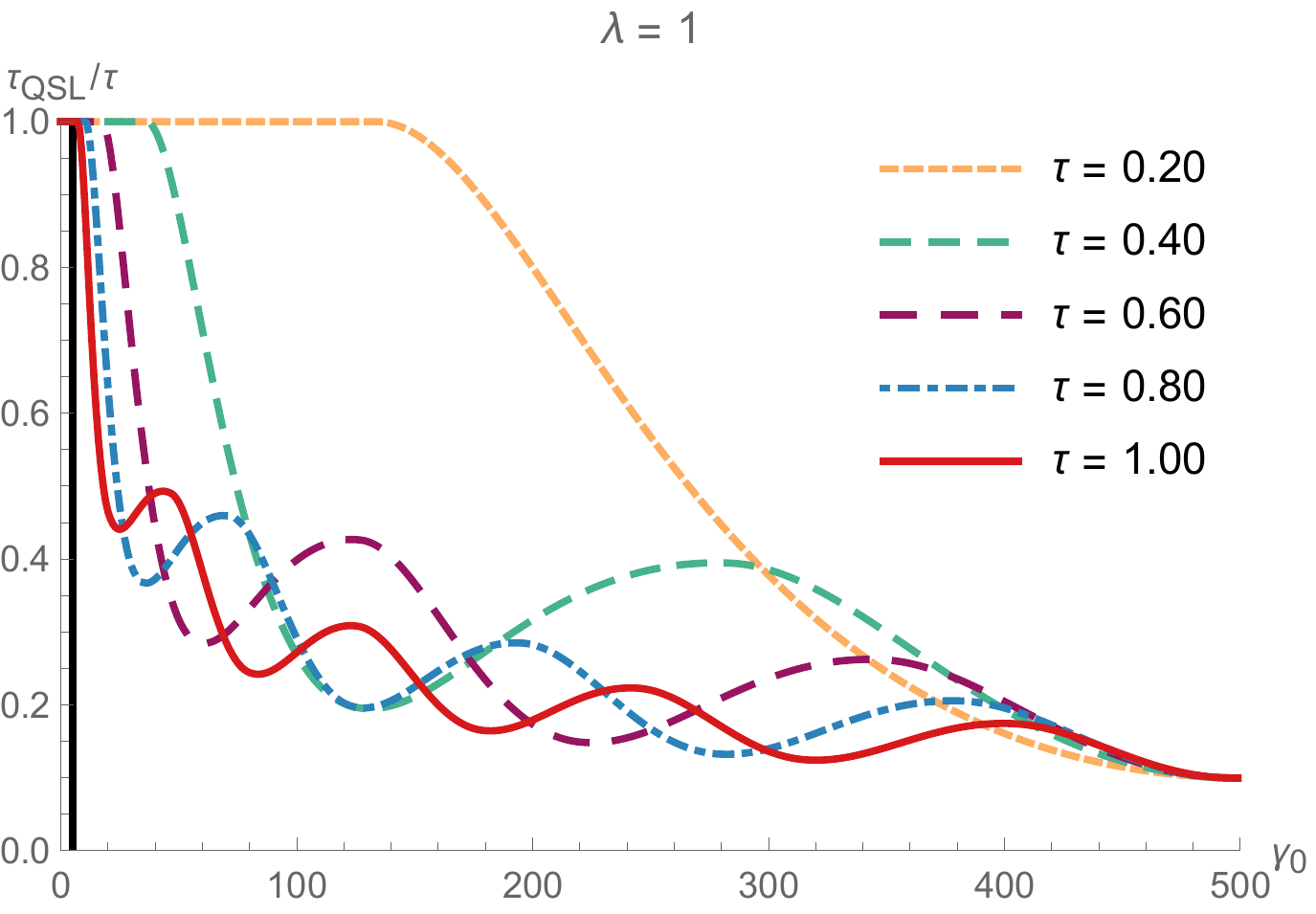}
\caption{The ratio $\tau_{QSL}/\tau$ as a function of the coupling constant $\gamma_0$, for different choices of $\tau$. The black vertical line is the critical value $\gamma_{0}^{crit}$ of $\gamma_0$.  We see, that the choice of $\tau$ affects the QSL in a non-monotonic way.
}\label{fig:difftaus}
\end{figure}

In Fig. \ref{fig:difftaus} we show the bound $\tau_{QSL}/\tau$ as a function of the coupling constant $\gamma_0$, for different choices of $\tau$. 
It is immediate to see that the QSL depends noticeably on the chosen evolution time on short intervals and that the QSL as a function of $\tau$ is not monotonic. The plateau of $\tau_{QSL}/\tau = 1$ in the non-Markovian regime of $\gamma_0 >\gamma_{0}^{crit}$ space is explained by the dynamics and the direct dependence of the BLP-measure and the $\tau_{QSL}/\tau$ in Eq.~\eqref{eq:BLP_form}: if the time interval is chosen so short that the dynamics exhibit no recoherence, the BLP-measure is zero, and thus $\tau_{QSL}/\tau = 1$.

\section{Connection between BLP non-Markovianity and QSL}\label{sec:blp_and_qsl}

As seen in Sec.~\ref{sec:jaynes_cummings}, there exists a connection between the values of the BLP non-Markovianity and QSL for the Jaynes-Cummings model: $\tau_{QSL}/\tau$ is a simple function of the BLP measure and $\tau_{QSL}/\tau = 1$ if and only if the dynamics is BLP Markovian. To generalize this result to other dynamical maps, we first solve the general requirements for an optimal QSL evolution.

We can analytically solve the optimal initial states, leading to $\tau_{QSL}/\tau = 1$. Trivially, for a pure initial state $\rho_0 = \vert \psi_0 \rangle \langle \psi_0 \vert$, we have $\bra{\psi_0}\rho_{0}\ket{\psi_0} = 1$, and thus 
$1 - \bra{\psi_0}\rho_{\tau = 0}\ket{\psi_0} = 0$. The QSL is reached for all $\tau'\in[0,\tau)$ if and only if,
\begin{align}
&\frac{ \tau_{QSL} }{ \tau' } 
= \frac{ 1 - \bra{\psi_0}\rho_{\tau'}\ket{\psi_0} }{ \int_0^{\tau'} || L_t(\rho_t) ||_{op} dt } 
= 1 ~~~~~~~~~~~~~~~\forall\tau'\in[0,\tau)\\
\Leftrightarrow &1 - \bra{\psi_0}\rho_{\tau'}\ket{\psi_0} 
= \int_0^{\tau'} || L_t(\rho_t) ||_{op} dt ~~~~~~~\forall\tau'\in[0,\tau)\\
\Rightarrow & -\frac{d}{d\tau'} \bra{\psi_0}\rho_{\tau'}\ket{\psi_0} 
= || L_{\tau'}(\rho_{\tau'}) ||_{op}~~~~~~~~~~\forall\tau'\in[0,\tau) \label{eq:optcond1} \\
\Rightarrow & 1 - \bra{\psi_0}\rho_{\tau'}\ket{\psi_0} = \int_0^{\tau'} || L_t(\rho_t) ||_{op} dt~~~~~~~\forall\tau'\in[0,\tau)
\end{align}
Since these equations form an equivalent chain, it suffices to study when the simpler condition \eqref{eq:optcond1} is satisfied.

By calculating the singular values and using the non-negativity of the operator norm, we see that for a qubit system, Eq.~\eqref{eq:optcond1} is equivalent to
\begin{align}
\bra{\psi_0} \dot{\rho}_\tau \ket{\psi_0^{\perp}} &= 0 ~~ \label{eq:cond_1}\\
\land ~~ \bra{\psi_0} \dot{\rho}_\tau \ket{\psi_0} &\leq 0 \label{eq:cond_2}\,,
\end{align}
where $\psi_0^\perp$ denotes the state orthogonal to $\psi_0$.

To further study the qubit case, we write the general Bloch vector dynamics $\mathbf{r}(t)$ as
\begin{equation}\label{eq:Bloch_map}
\mathbf{r}(t) = A(t) \mathbf{r}(0) + \mathbf{s}(t)
\end{equation}
with
\begin{align}
A(t) &= \left(
\begin{array}{ccc}
a_{11}(t) &a_{12}(t) &a_{13}(t) \\
a_{21}(t) &a_{22}(t) &a_{23}(t) \\
a_{31}(t) &a_{32}(t) &g(t)
\end{array}
\right)
\text{~(deformation)}  \,, \\
\mathbf{r}(0) &= \left(
\begin{array}{c}
x(0) \\
y(0) \\
z(0)
\end{array}
\right)
\text{~~~~~~~~~~~~~~~~~~~~~~(initial state)}  \,, \\
\mathbf{s}(t) &= \left(
\begin{array}{c}
s_1(t) \\
s_2(t) \\
h(t)
\end{array}
\right)
\text{~~~~~~~~~~~~~~~~~~~~~\,(translation)}  \,.
\end{align}
We fix the basis, so that $\lbrace \ket{\psi_0^-}, \ket{\psi_0^+} \rbrace$, corresponding to $r(0) = (0,0,\pm 1)^T$, is the optimal pair of initial states for the BLP-measure.  The results of \cite{wissmann2012} guarantee that the optimal pair of initial qubit states maximizing the BLP measure can always be chosen as an orthogonal pair of pure states. Based on Eqs.~\eqref{eq:cond_1} and \eqref{eq:cond_2}, we can study the relationship between BLP-measure and $\tau_{QSL}/\tau$ more generally.
\begin{figure}
\begin{minipage}[h!]{1\textwidth}
\centering
\includegraphics[width=0.603\textwidth]{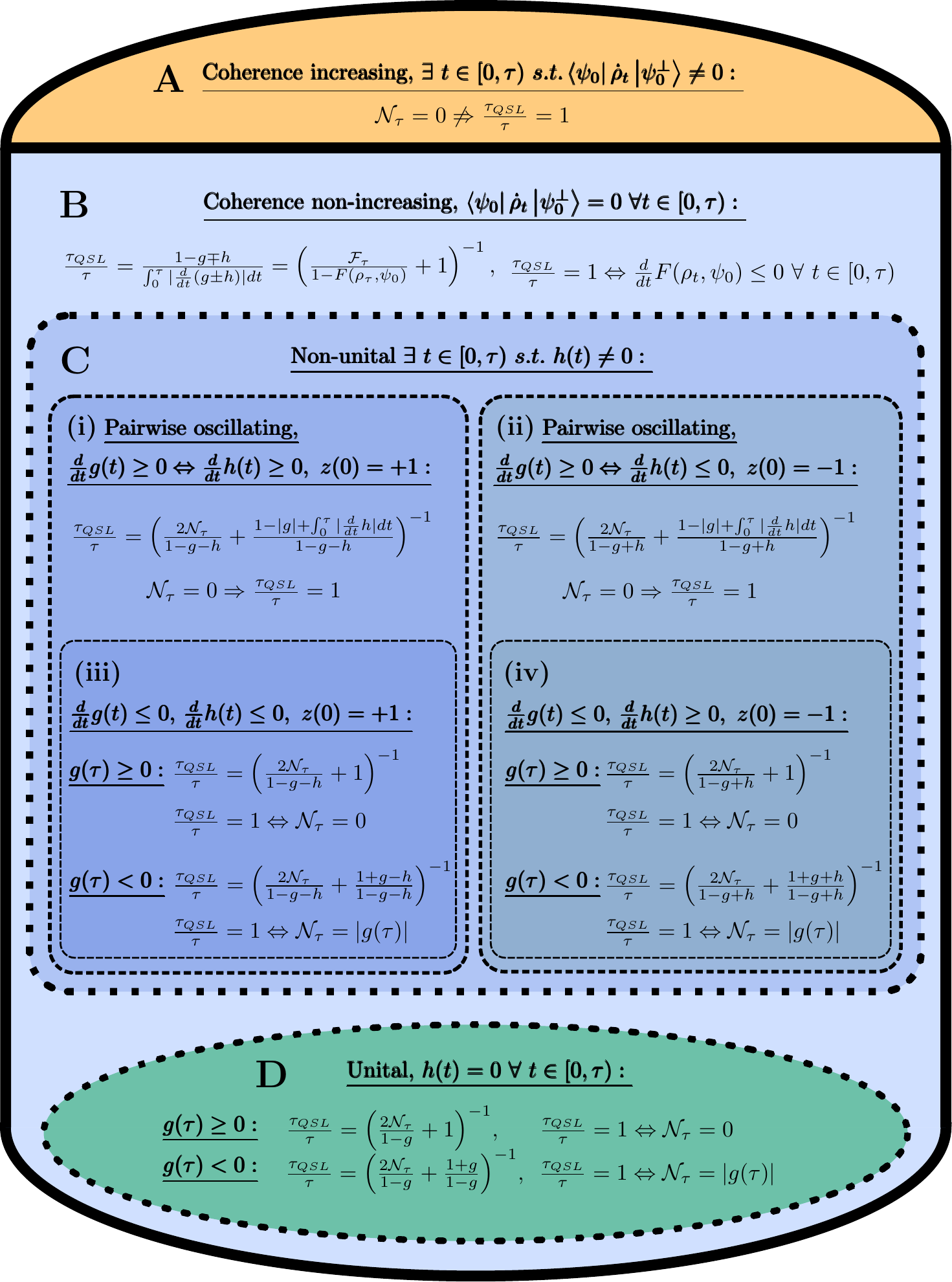}
\caption{
\footnotesize{
\textbf{Summary of QSL 
as a function of BLP non-Markovianity and fidelity 
for all CPTP one-qubit dynamical maps:} Each set of dynamical maps \textbf{A}--\textbf{D} is characterized by the underlined condition(s) in bold text. Each condition has to be satisfied for all times $t\in[0,\tau)$, unless stated otherwise (\textbf{A} and \textbf{C}). In \textbf{B} the upper and lower signs in $\pm$ and $\mp$ correspond to the choices of initial states $z(0) = +1$ and $z(0) = -1$, respectively. The subset inherits the condition(s) of its superset. The inclusion hierarchy of the sets is $\textbf{B} = \textbf{C} \mathbin{\mathaccent\cdot\cup} \textbf{D}$,   $\textbf{(iii)} \subset\textbf{(i)} \subset \textbf{C}$,  $\textbf{(iv)} \subset \textbf{(ii)} \subset \textbf{C}$. For brevity, we have omitted the explicit time dependence of $g = g(\tau)$, $g = g(t)$, $h = h(\tau)$, $h = h(t)$. $\mathcal{F}_\tau = \int_{\frac{d}{dt} [g(t) \pm h(t)] > 0 }^{t \in (0,\tau)}  \frac{d}{dt} \left[ g(t) \pm h(t) \right] dt$ is the sum of temporal revivals of fidelity between the initial and evolved states $\psi_0$ and $\rho_\tau$, and $\mathcal{N}_\tau = \mathcal{N}(\Phi_\tau)$ is the BLP non-Markovianity of the dynamical map. 
   In \textbf{A}, the QSL bound is not always reached with the optimal initial states of the BLP measure even though the dynamics would be Markovian
   , so the 
   BLP measure is not critical for tightness of the bound. 
   For \textbf{B}, $\tau_{QSL}/\tau$ can be expressed in terms of the fidelity between the initial state $\psi_0$ and the evolved state $\rho_\tau$ and its total temporal revivals $\mathcal{F}_\tau$. 
   After the first revival of fidelity, $\tau_{QSL}/\tau$ becomes a monotonically decreasing function of $F(\rho_\tau,\psi_0)$. 
   In \textbf{C (i)} and \textbf{(ii)}, we see how $\tau_{QSL}/\tau$ depends explicitly on the non-Markovianity: BLP Markovianity implies tightness of QSL bound and if $g(\tau) \geq 0$, we get $\tau_{QSL}/\tau = 1 \Leftrightarrow \mathcal{N}(\Phi_\tau) = 0$. For their subsets \textbf{C (iii)} and \textbf{(iv)}, the result of $g(\tau) \geq 0$ still holds, but surprisingly when $g(\tau) < 0$ the condition is expanded into $\tau_{QSL}/\tau = 1 \Leftrightarrow \mathcal{N}(\Phi_\tau) = |g(\tau)|$. Thus, tightness of the QSL bound does not guarantee BLP Markovianity in cases where the behavior of both $g(t)$ and $h(t)$ is monotonic $\forall~t\in[0,\tau)$. As special cases, \textbf{C (i)} and \textbf{C (ii)} contain the Jaynes Cummings model and the whole set of commutative phase-covariant dynamics, respectively. 
   \textbf{D} is the set of all CPTP unital one-qubit-maps satisfying the condition of \textbf{B}. As in the case of \textbf{C (iii)} and \textbf{(iv)} we see that $\tau_{QSL}/\tau = 1 \Leftrightarrow \mathcal{N}(\Phi_\tau) = 0$ when $g(\tau) \geq 0$ and $\tau_{QSL}/\tau = 1 \Leftrightarrow \mathcal{N}(\Phi_\tau) = |g(\tau)|$ if $g(\tau) < 0$.
   }
   }
\label{fig:classes_of_dynamics}
\end{minipage}
\end{figure}
In the following, we divide the set of all one-qubit dynamical maps into subsets illustrated in Fig.~\ref{fig:classes_of_dynamics} and analyze the connection between BLP non-Markovianity and tightness of the QSL bound.

\subsection{Coherence-increasing and coherence non-increasing maps}

If the coherences between $\ket{\psi_0^+}$ and $\ket{\psi_0^-}$ increase for $t\in[0,\tau)$, as in Fig.~\ref{fig:classes_of_dynamics} \textbf{A}, Eq.~\eqref{eq:cond_1} is violated. If the violation occurs at $t = 0$, we have $\tau_{QSL}/\tau < 1$ for all times $\tau \geq 0$. Furthermore, positivity of the dynamical map requires that the BLP non-Markovian behaviour does not begin at $\tau = 0$, so $\tau_{QSL}/\tau < 1$ already in the Markovian region, and thus $\tau_{QSL}/\tau$ does not critically depend on the BLP-measure. The same reasoning holds for all cases where the coherences increase at any time before the first non-Markovian effects take place.

For the initial state given by the Bloch vector $r(0) = (0,0,\pm 1)^T$, the dynamical map does not increase the coherences between $\ket{\psi_0^+}$ and $\ket{\psi_0^-}$, and Eq.~\eqref{eq:cond_1} is satisfied if and only if $x(t) = y(t) = 0$ and $z(t) = g(t) z(0) + h(t)$. This class of dynamics corresponds to Fig.~\ref{fig:classes_of_dynamics} \textbf{B}. For such dynamics, $\tau_{QSL}/\tau$ can be written as
\begin{equation}\label{eq:general_ratio_form}
\frac{\tau_{QSL}}{\tau} = \frac{1 - g(\tau) \mp h(\tau)}{ \int_0^\tau |\frac{d}{dt}(g(t) \pm h(t))| dt} = \left( \frac{\mathcal{F}_\tau}{1-F(\rho_\tau, \psi_0^\pm)} + 1 \right)^{-1}\,,
\end{equation}
where $F(\rho_\tau,\psi_0^\pm) = \bra{\psi_0^\pm} \rho_\tau \ket{\psi_0^\pm}$ is the fidelity between the initial state and the evolved state at time $\tau$ and $\mathcal{F}_\tau = \int_{\frac{d}{dt} [g(t) \pm h(t)] > 0 }^{t \in (0,\tau)} \frac{d}{dt} \left[ g(t) \pm h(t) \right] dt$ is the sum of temporal revivals of $F(\rho_t,\psi_0^\pm)$. We see directly from Eq.~\eqref{eq:general_ratio_form} that $\tau_{QSL}/\tau = 1 \Leftrightarrow \mathcal{F}_\tau = 0$, so oscillations of the fidelity are necessary to decrease $\tau_{QSL}/\tau$ and after the first oscillation $\tau_{QSL}/\tau$ is always smaller than 1. When $\mathcal{F}_\tau > 0$, $\tau_{QSL}/\tau$ is a monotonically decreasing function of $F(\rho_\tau, \psi_0^\pm)$: When the fidelity between the initial and evolved states increases, $\tau_{QSL}/\tau$ decreases and vice versa. Also, $\tau_{QSL}/\tau = 0$ if and only if there has been increase of the fidelity and $\rho_\tau = \psi_0$. 
As we will see in the following, $\tau_{QSL}/\tau = 1$ is not equivalent to $\mathcal{N}(\Phi_\tau) = 0$, due to $h(t)$ dependence of $\mathcal{F}_\tau$, and in some cases, $\mathcal{N}(\Phi_\tau) > 0$ does not lead to $\tau_{QSL}/\tau < 1$.

In the following subsections, we will study some relevant subclasses of the coherence non-increasing maps and derive the explicit dependency between $\tau_{QSL}/\tau$ and the BLP measure.
\subsection{Pairwise oscillating translation and deformation}
Let us concentrate here on Fig.~\ref{fig:classes_of_dynamics} \textbf{C (i)} (and \textbf{(ii)}), where the translation always increases (or decreases) exactly when the deformation increases and vice versa. First, assuming $\tfrac{d}{dt} g(t) \geq 0 ~\Leftrightarrow \tfrac{d}{dt} h(t) \geq 0~\forall t  \in [0,\tau)$ and choosing $z(0) = +1$ in Fig.~\ref{fig:classes_of_dynamics} \textbf{C (i)}, Eq.~\eqref{eq:general_ratio_form} becomes
\begin{equation}\label{eq:32}
\begin{split}
\frac{\tau_{QSL}}{\tau} = \left( \frac{2\mathcal{N}(\Phi_\tau)}{1- g(\tau) - h(\tau)} + \frac{1 - |g(\tau)| + \int_0^\tau | \frac{d}{dt} h(t) | dt}{1- g(\tau) - h(\tau)} \right)^{-1} \,.
\end{split}
\end{equation}
Here
\begin{equation}
\mathcal{N}(\Phi_\tau) = \int_{\frac{d}{dt}|g(t)| > 0} \frac{d}{dt} |g(t)|  dt \,,
\end{equation}
meaning that the BLP measure is independent of the translation $h(t)$, unlike the QSL. When $\mathcal{N}(\Phi_\tau) = 0$, we have $g(t) \geq 0$, $h(t) \leq 0$, and $\tfrac{d}{dt} h(t) \leq 0~\forall t \in (0,\tau)$. In the case of Eq.~\eqref{eq:32} this means $\tau_{QSL}/\tau = 1$, even if $h(t) \neq 0$. Thus in this situation $\tau_{QSL}/\tau < 1$ only if the non-Markovian effects have kicked in. If in addition $g(\tau) \ge 0$, we note that $\tau_{QSL}/\tau = 1 \Leftrightarrow \mathcal{N}(\Phi_\tau) = 0$.

In the special case Fig.~\ref{fig:classes_of_dynamics} \textbf{C (iii)}, when $\tfrac{d}{dt}g(t) \leq 0,~\tfrac{d}{dt}h(t) \leq 0~\forall t \in [0,\tau)$, the BLP dependency can be broken into two cases based on the sign of $g(\tau)$: If $g(\tau) \geq 0$, the QSL can be written as
\begin{equation}
\frac{\tau_{QSL}}{\tau} = \left( \frac{2 \mathcal{N}(\Phi_\tau)}{1 - g(\tau) - h(\tau)} + 1 \right)^{-1} \,,
\end{equation}
and thus $\tau_{QSL}/\tau = 1 \Leftrightarrow \mathcal{N}(\Phi_\tau) = 0$. Let us now consider the situation where $g(t)$ is a continuous function which decreases monotonically until $t'$, so that $g(t') = 0$. Now $\mathcal{N}(\Phi_{t'}) = 0$ as $|g(t)|$ is also monotonic in the interval $[0,t']$. As $g(t)$ continues to decrease monotonically until $\tau$, the QSL becomes
\begin{equation}
\frac{\tau_{QSL}}{\tau} = \left( \frac{2 \mathcal{N}(\Phi_\tau)}{1 - g(\tau) - h(\tau)} + \frac{1+g(\tau)-h(\tau)}{1-g(\tau)-h(\tau)} \right)^{-1} \,,
\end{equation}
since $g(\tau) <0$, and we see that $\tau_{QSL}/\tau = 1 \Leftrightarrow \mathcal{N}(\Phi_\tau) = |g(\tau)| $. Thus, in this case we have optimal evolution even if the dynamics is non-Markovian.

Similarly, assuming $\tfrac{d}{dt} g(t) \geq 0 ~\Leftrightarrow \tfrac{d}{dt} h(t) \leq 0$, $\forall t \in [ 0,\tau)$ and choosing $z(0) = -1$ in Fig.~\ref{fig:classes_of_dynamics} \textbf{C (ii)}, Eq.~\eqref{eq:general_ratio_form} yields to
\begin{equation}\label{eq:33}
\frac{\tau_{QSL}}{\tau} = \left( \frac{2\mathcal{N}(\Phi_\tau)}{1- g(\tau) + h(\tau)} + \frac{1 - |g(\tau)| + \int_0^\tau | \frac{d}{dt} h(t) | dt}{1- g(\tau) + h(\tau)} \right)^{-1} \,,
\end{equation}
and we obtain the same dependency between tightness of the QSL bound and non-Markovianity (see Fig.~\ref{fig:classes_of_dynamics} \textbf{C (ii)} and \textbf{(iv)}).

\subsection{Unital maps}
The considerations made above hold for generic translations, including the non-unital cases $h(t) \neq 0$. Now, we restrict to the unital maps in Fig.~\ref{fig:classes_of_dynamics} \textbf{D}, characterized by $h(t) = 0~\forall t \geq 0$, for which Eq.~\eqref{eq:general_ratio_form} becomes
\begin{equation}\label{eq:h_zero_qsl}
\frac{\tau_{QSL}}{\tau} = \left( \frac{2 \mathcal{N}(\Phi_\tau)}{1-g(\tau)} + \frac{1 - |g(\tau)|}{1-g(\tau)} \right)^{-1} \,.
\end{equation}
We note that Eq.~\eqref{eq:h_zero_qsl} can be written as
\begin{equation}
\frac{\tau_{QSL}}{\tau} = \left( \frac{2 \mathcal{N}(\Phi_\tau)}{1-g(\tau)} + 1 \right)^{-1} \,,
\end{equation}
if and only if $g(\tau) \geq 0$. This means exactly the same dependence on the BLP-measure as in the case of Eq.~\eqref{eq:BLP_form} if $g(\tau) \geq 0$. If instead $g(\tau)<0$, we can write the QSL as
\begin{equation}
\frac{\tau_{QSL}}{\tau} = \left( \frac{2 \mathcal{N}(\Phi_\tau)}{1 - g(\tau)} + \frac{1+g(\tau)}{1-g(\tau)} \right)^{-1} \,,
\end{equation}
which leads to $\tau_{QSL}/\tau = 1 \Leftrightarrow \mathcal{N}(\Phi_\tau) = |g(\tau)|$, implying optimal evolution for non-Markovian dynamics. This means that if $\mathcal{N}(\Phi_t) = 0$ still when $g(t)$ becomes negative, $\tau_{QSL}/\tau$ begins to decrease exactly when the non-Markovian behavior ends, which is the opposite of what happens in the Jaynes-Cummings model.

In the above considerations, we assumed that $\psi_0^+$ and $\psi_0^-$ are the optimal initial states maximizing the BLP measure. But even if the initial states were not the optimal pair, all the above analysis would still hold. The only exception would be that $\mathcal{N}(\Phi_\tau)$ would just quantify information backflow in terms of increased distinguishability of these sub-optimal states, thus losing the exact interpretation of BLP measure of non-Markovianity. We will conclude this section with an example class of dynamics belonging to Fig.~\ref{fig:classes_of_dynamics} \textbf{C (ii)}.

\subsection{Example: phase-covariant commutative dynamics}
As an example, we use the phase-covariant system of Eq.~\eqref{eq:pahsecovariant_me} which does not increase coherences between $\ket{0}$ and $\ket{1}$ \cite{teittinen2018}. For the commutative class of phase-covariant dynamics, that is when $\gamma_1(t) = \gamma(t)$ and $\gamma_2(t) = \kappa \gamma(t)$, with $0 \leq \kappa \leq 1$, the functions $g(t)$ and $h(t)$ have the form
\begin{align}
g(t) = e^{-\Gamma(t)} \,, ~~~~
h(t) = \frac{1 - \kappa}{1 + \kappa} \left( 1 - e^{-\Gamma(t)} \right) \,,
\end{align}
where $\Gamma(t) =\tfrac{\kappa + 1}{2} \int_0^t \gamma(t') dt'$. \footnote{To be precise, the commutative class contains also the cases where $\gamma_1(t) = \kappa \gamma(t)$ and $\gamma_2(t) = \gamma(t)$, which belong to \textbf{C (i)} in Fig.~\ref{fig:classes_of_dynamics}. If we choose $\kappa = 1$, that is when $\gamma_1(t) = \gamma_2(t)$, the dynamics is unital and belongs to \textbf{D} in Fig.~\ref{fig:classes_of_dynamics}.} Since $\tfrac{d}{dt}g(t) \geq 0 \Leftrightarrow \tfrac{d}{dt}h(t) \leq 0$, we can write Eq.~\eqref{eq:33} for this system as
\begin{equation}
\frac{\tau_{QSL}}{\tau} = \left(
\frac{2 \mathcal{N}(\Phi^{PC}_\tau)}{\frac{2}{\kappa +1}(1 - e^{-\Gamma(\tau)})} + \frac{1 - e^{-\Gamma(\tau)} + \frac{1-\kappa}{2}\int_0^\tau |\gamma(t)|e^{-\Gamma(\tau)}dt}{\frac{2}{\kappa +1}(1 - e^{-\Gamma(\tau)})} \right)^{-1} \,,
\end{equation}
with
\begin{align}
\mathcal{N}(\Phi_\tau^{PC}) 
&= \int_{\gamma(t) < 0} - \frac{1+ \kappa}{2} \gamma(t) e^{-\Gamma(t)} dt \,.
\end{align}
If $\gamma(t) \geq 0, ~\forall t \in [0,\tau)$, then $\mathcal{N}(\Phi_\tau^{PC}) = 0$ and $\tau_{QSL}/ \tau = 1$.
We also notice, that in this case, $|\tfrac{d}{dt}g(t)| \geq |\tfrac{d}{dt}h(t)|~\forall~t\ge 0$, and thus $\tfrac{d}{dt} (g(t) + h(t))$ is dominated by $\tfrac{d}{dt}g(t)$. Since the derivatives of $g(t)$ and $h(t)$ change sign at the same time, the sign of $\tfrac{d}{dt} (g(t) + h(t))$ is always the sign of $\tfrac{d}{dt} g(t)$. As a consequence, $\tau_{QSL}/\tau = 1 \Leftrightarrow \mathcal{N}(\Phi_\tau) = 0$ for both choices of initial state $\psi_0^+$ and $\psi_0^-$.

\section{Initial state dependence of QSL for Markovian master equations}\label{sec:markovian_mes}
To continue the generalization of our results, we now take a complementary perspective: instead of looking at the connection between the values of the BLP non-Markovianity measure and the QSL, we focus on the families of initial states leading to saturation of the quantum speed limit time in the Markovian case. The results of Deffner and Lutz suggest that Markovian dynamics results always in optimal time, that is $\tau_{QSL}/\tau = 1$, for the Jaynes-Cummings system. For some pure initial states this is true, but not for all, when looking at more general Markovian master equations.

\subsection{Phase-covariant}

Here we study the dynamics described by the master equation of Eq.~\eqref{eq:pahsecovariant_me}, with $\gamma_1(t) = \gamma_1$, $\gamma_2(t) = \gamma_2$, and $\gamma_3(t) = \gamma_3$, $\forall t$, where $\gamma_1,\gamma_2,\gamma_3 \geq 0$. We notice that the phase difference between $\ket{0}$ and $\ket{1}$ does not have any significant role, in the phase covariant master equation, with respect to the QSL. Thus we parametrize the initial state as $\rho_0 = \ket{\psi_0}\bra{\psi_0}$, where $\ket{\psi_0} = \sqrt{a} \ket{1} + \sqrt{1-a} \ket{0}$. Now, we characterize the set of initial states leading to $\tau_{QSL}/\tau = 1$ for all $\tau \ge 0$ by using \eqref{eq:optcond1}, which becomes:
\begin{equation}\label{eq:optimal_condition_pc}
\begin{split}
\frac{1}{16}(a-1)a e^{-(\gamma_1 + \gamma_2 + 2\gamma_3)t} &\left[ -4e^{\gamma_3 t} ((a-1)\gamma_1 + a\gamma_2) \right. \\ &\left. - (1-2a)e^{(\gamma_1 + \gamma_2)t/4} (\gamma_1 + \gamma_2 +4 \gamma_3) \right] = 0 \,.
\end{split}
\end{equation}
We see, that now we have $\tau_{QSL} /\tau = 1~\forall~\tau\ge 0$ if and only if $a = 0$ or $a = 1$. We emphasize, that these are not stationary states, but initial states that always evolve with the optimal QSL time. If we restrict to the unital case $\gamma_1 = \gamma_2$ with $\omega = 0$, also the initial state $a = 1/2$ leads to $\tau_{QSL} /\tau = 1~\forall~\tau\ge 0$. We note that $\tau_{QSL}/\tau = 1 , ~\forall a \in [0,1]$ if and only if $\gamma_1 = \gamma_2 = 2 \gamma_3$. In this case, the dynamical map is of the depolarizing form
\begin{equation}
\rho_t = (1 - p(t)) \rho_0 + p(t) \frac{1}{2} \mathbb{I} \,,
\end{equation}
where $p(t) \in [0,1]$, with $p(0) = 0$.

\begin{figure}\centering
\includegraphics[width=0.5\textwidth]{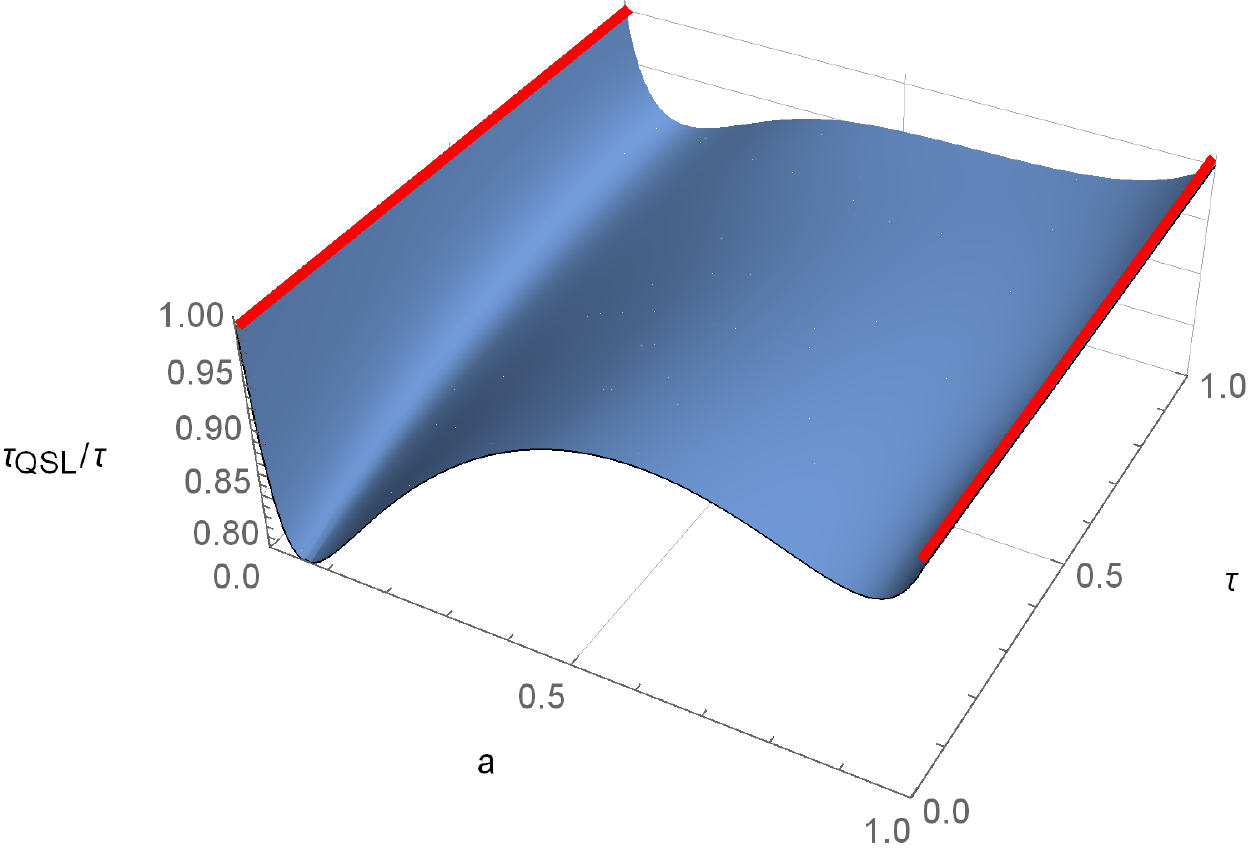}
\caption{$\tau_{QSL}/\tau$ for the phase covariant channel with $\gamma_1 = 1$, $\gamma_2 = 2$, $\gamma_3 = 3$, and $\omega = 0$ as a function of the evolution time $\tau \in [0,1]$ and the initial state parameter $a \in [0,1]$. Optimal initial states at $a=0$ and $a=1$. The local maximum near $a=1/2$ is affected by the balance between $\gamma_1$ and $\gamma_2$, in this case, $\gamma_2 > \gamma_1$ placing the maximum closer to $a=1$. If $\gamma_1 = \gamma_2$, this coincides with the Pauli channel and the optimal states are found at $a=0$, $a=1$, and $a=1/2$. Generally the value of $\tau_{QSL}/\tau$ is not constant w.r.t. $\tau$ in regions where $\tau_{QSL}/\tau<1$. The red highlights represent the points where $\tau_{QSL}/\tau = 1$.}\label{fig:QSL_a_optimize_phasecovariant}
\end{figure}

Fig. \ref{fig:QSL_a_optimize_phasecovariant} shows the initial state and $\tau$ dependence of the phase-covariant master equation for $\gamma_1 = 1,\, \gamma_2 = 2,\, \gamma_3 = 3$. Again, we see, that the optimal points are found at $a=0$ and $a=1$, that is diagonal pure states w.r.t.~the $\{\ket{0},\ket{1}\}$ basis, while all other states fail to reach the limit.

\subsection{Pauli channel}

Now, we consider the system described by the master equation of Eq.~\eqref{eq:pauli_me}, with $\gamma_1(t) = \gamma_1$, $\gamma_2(t) = \gamma_2$, and $\gamma_3(t) = \gamma_3$, $\forall t$, where $\gamma_1,\gamma_2,\gamma_3 \geq 0$. The unital case of the phase-covariant master equation, that is when $\gamma_1 = \gamma_2$, coincides with the Pauli channel, with the same decay rates. However, the general Pauli channel covers a larger set of dynamics than the unital phase-covariant, such as bit-flip and bit-phase-flip channels.

\begin{figure}\centering
\includegraphics[width=0.5\textwidth]{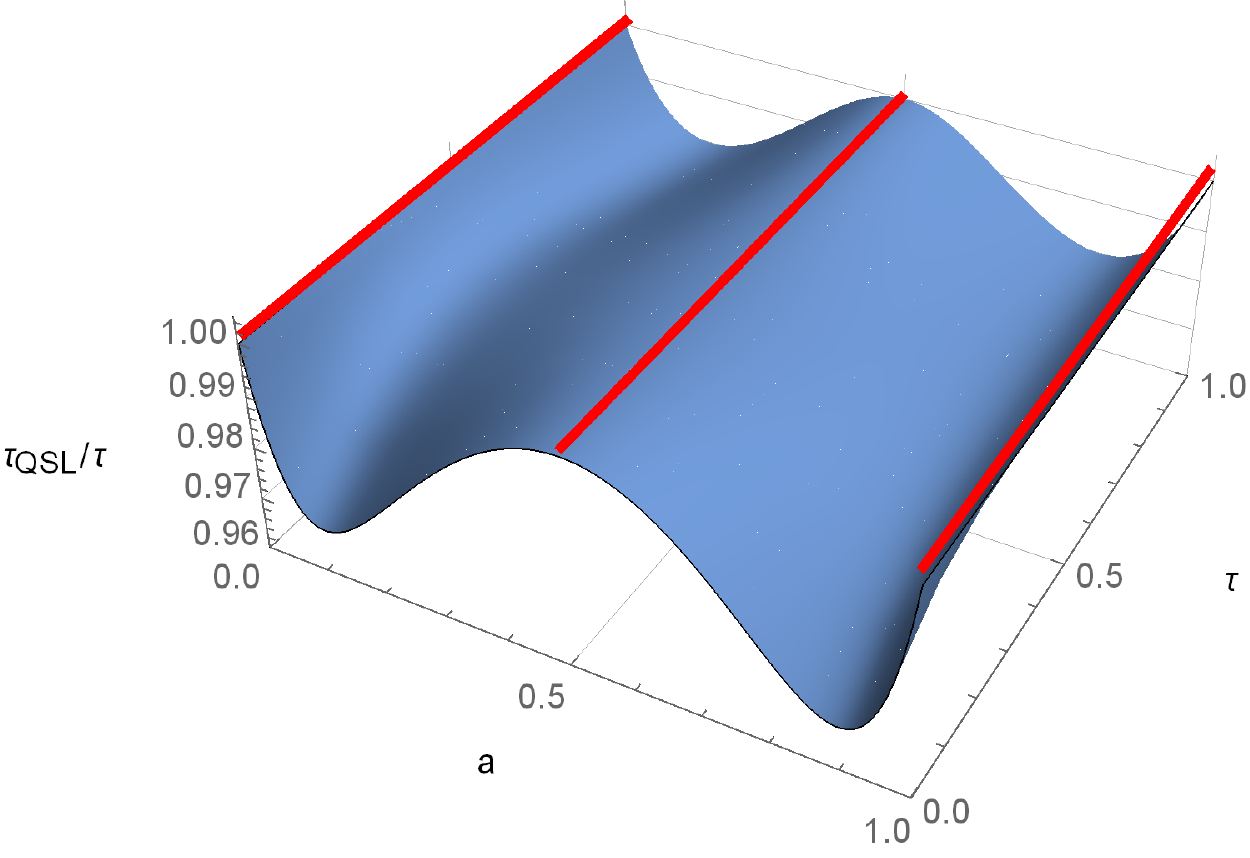}
\caption{$\tau_{QSL}/\tau$ for the Pauli channel with $\gamma_1 = 1$, $\gamma_2 = 2$, and $\gamma_3 = 3$ as a function of the evolution time $\tau \in [0,1]$ and the initial state parameter $a \in [0,1]$. Optimal choices at $a=0$, $a=1$, and $a = 1/2$. Generally the value of $\tau_{QSL}/\tau$ is not constant w.r.t.~$\tau$ in regions where $\tau_{QSL}/\tau<1$. The red highlights represent the points where $\tau_{QSL}/\tau = 1$.}\label{fig:QSL_a_optimize_pauli}
\end{figure}

As for the phase-covariant model, we can analytically derive the optimal states using Eq.~\eqref{eq:optcond1}. The resulting condition is:
\begin{equation}
(1 - 2 a)^2 (a-1) a e^{-2 t (\gamma_1 + \gamma_2 + \gamma_3)} \big[ e^{ \gamma_3 t} (\gamma_1 + \gamma_2) - e^{ \gamma_1 t} (\gamma_2 + \gamma_3) \big]^2 = 0\,.
\end{equation}
We see, that the QSL is reached $\forall t \geq 0$, with $a=0$, $a=1$, and $a=1/2$. Similarly to the case of phase-covariant master equation, by choosing $\gamma_1 = \gamma_3$, $\tau_{QSL}/\tau = 1 , ~\forall a \in [0,1]$. By extending the initial states to cover all pure states, $\ket{\psi_0} = a \ket{0} + e^{i \theta} \sqrt{1-a} \ket{0}$, we get $\tau_{QSL}/\tau = 1$ for all $\theta \in [0,2\pi]$ and $a \in [0,1]$, when $\gamma_1 = \gamma_2 = \gamma_3$. Fig. \ref{fig:QSL_a_optimize_pauli} shows the initial state dependence of QSL for Pauli channel with $\gamma_1 = 1,\, \gamma_2 = 2,\, \gamma_3 = 3.$

\subsection{Eternal non-Markovianity}

The eternal non-Markovianity model is interesting in this context since it is always completely positive and non-CP-divisible ($\gamma_3(t) < 0~\forall t > 0$), but at the same time BLP-Markovian. The eternally non-Markovian master equation has the form \cite{hall2014}:
\begin{equation}
L_t\rho_t = \frac{1}{2} \left( \sigma_1 \rho_t \sigma_1 - \rho \right) + \frac{1}{2} \left( \sigma_2 \rho_t \sigma_2 - \rho \right) - \frac{\tanh(t)}{2} \left( \sigma_3 \rho_t \sigma_3 - \rho \right)
\end{equation}
The condition for reaching the QSL in this case is given by
\begin{equation}
(1-2a)^2 (a-1)a e^{-4t} = 0 \,,
\end{equation}
for which the solutions are $a=0$, $a=1$, and $a=1/2$. Since the eternally non-Markovian  model is a special case of the phase-covariant commutative master equation, with $\kappa = 1$ and translation h(t) = 0, we can compare the results of this analysis with the ones derived in Sec.~\ref{sec:blp_and_qsl}. We see, that the analytical results in Sec.~\ref{sec:blp_and_qsl} are in full agreement with this approach.

\begin{figure}\centering
\includegraphics[width=0.5\textwidth]{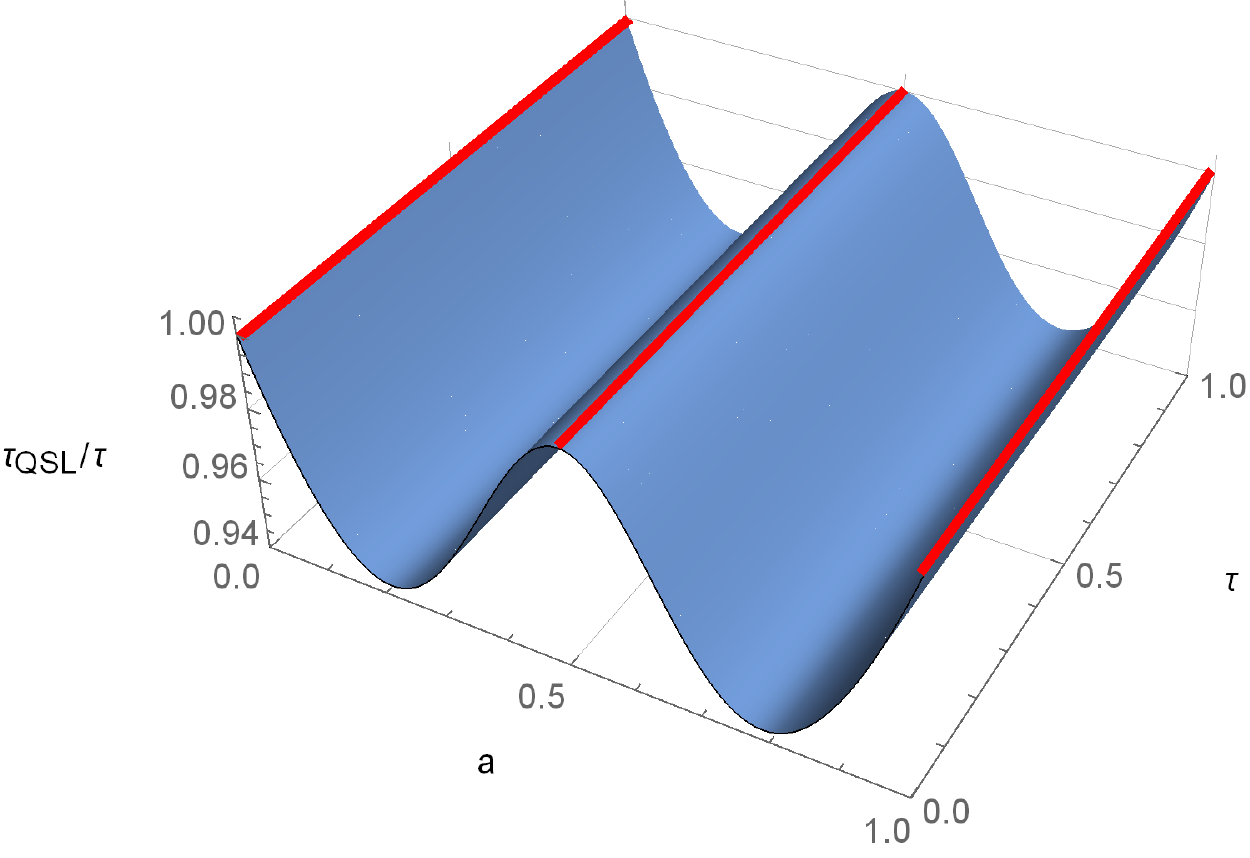}
\caption{$\tau_{QSL}/\tau$ for the eternal N-M channel as a function of the evolution time $\tau \in [0,1]$ and the initial state parameter $a \in [0,1]$. Optimal states at $a = 1$, $a = 0$, and $a = 1/2$. Despite being fully non-CP-divisible, the system has states for which $\tau_{QSL}/\tau = 1,~\forall~\tau \geq 0$. The red highlights represent the points where $\tau_{QSL}/\tau = 1$.}\label{fig:QSL_a_optimize_ENM}
\end{figure}

\section{The effect of Markovian-to-non-Markovian transition in QSL time}\label{sec:timedependent}

We now study the Markovian to non-Markovian transition using the results reported in Ref. \cite{teittinen2018}. We choose $\gamma_1(t)$, $\gamma_2(t)$, and $\gamma_3(t)$ and pinpoint the times at which a transition happens in the $\lbrace \gamma_3(t), \gamma'(t) \rbrace$-space, where $\gamma'(t) \equiv \gamma_1(t) + \gamma_2(t)$, without calculating explicitly any non-Markovianity measures. Thus, we avoid the initial state optimization required for the BLP-measure.

As an example, we will use the phase-covariant master equation, with the following decay rates: 
\begin{align}\label{tim-dep-model}
\begin{aligned}
\gamma_1 (t) = \gamma_2(t) = e^{-t/4}( 1 + \sin(t)) \,, ~~~~
\gamma_3 (t) = 2e^{-t/4} \cos(t), ~~~~ \omega(t) = 0 \,.
\end{aligned}
\end{align}
Since the master equation is in the Lindblad form and $\gamma_3(t)$ can have negatives values, we know that this dynamics is not CP-divisible, but is still CPTP according to the results of \cite{lankinen2016}.
The condition for optimal evolution from Eq.~\eqref{eq:optcond1} for this system is
\begin{equation}\label{e1cond}
\begin{split}
\frac{e_1(t)}{2}  \Bigg[ (1-2a)^2 f(t) -2 (a-1)& a e^{2} k(t) \\ &-\sqrt{(1-2a)^2 f(t)^2 - (a-1) a e^4 k(t)^2}\Bigg] = 0 \,,
\end{split}
\end{equation}
where $e_1(t)$, $f(t)$, and $k(t)$ are non-zero time dependent, but not $a$ dependent functions. Equation \eqref{e1cond} has solutions at $a=0$ and $a=1$. For the case $a=1/2$, the condition becomes
\begin{equation}
\begin{split}
\frac{e_2(t)}{4}  \Bigg[ 1 + 4 \cos(t) + \sin(t) - | 1 + 4\cos(t) + \sin(t) | \Bigg] = 0 \,,
\end{split}
\end{equation}
where $e_2(t)$ is a non-zero time-dependent, but not $a$ dependent function. We notice, that the condition is satisfied, when $1 + 4 \cos(t) + \sin(t) \geq 0$, but broken elsewhere. Thus, violation of $1 + 4 \cos(t) + \sin(t) \geq 0$ implies $\tau_{QSL}/\tau < 1$. According to \cite{teittinen2018}, this dynamical map is BLP non-Markovian if and only if $\gamma_1(t) + \gamma_2(t) + 4\gamma_3(t) < 0$ which in this case is equivalent to $1 + 4 \cos(t) + \sin(t) < 0$. So, we see that for $a = 1/2$ BLP non-Markovianity begins exactly at the same time as $\tau_{QSL}/\tau$ starts to decrease.

\begin{figure}[!t]
\centering
\includegraphics[width=.5\textwidth]{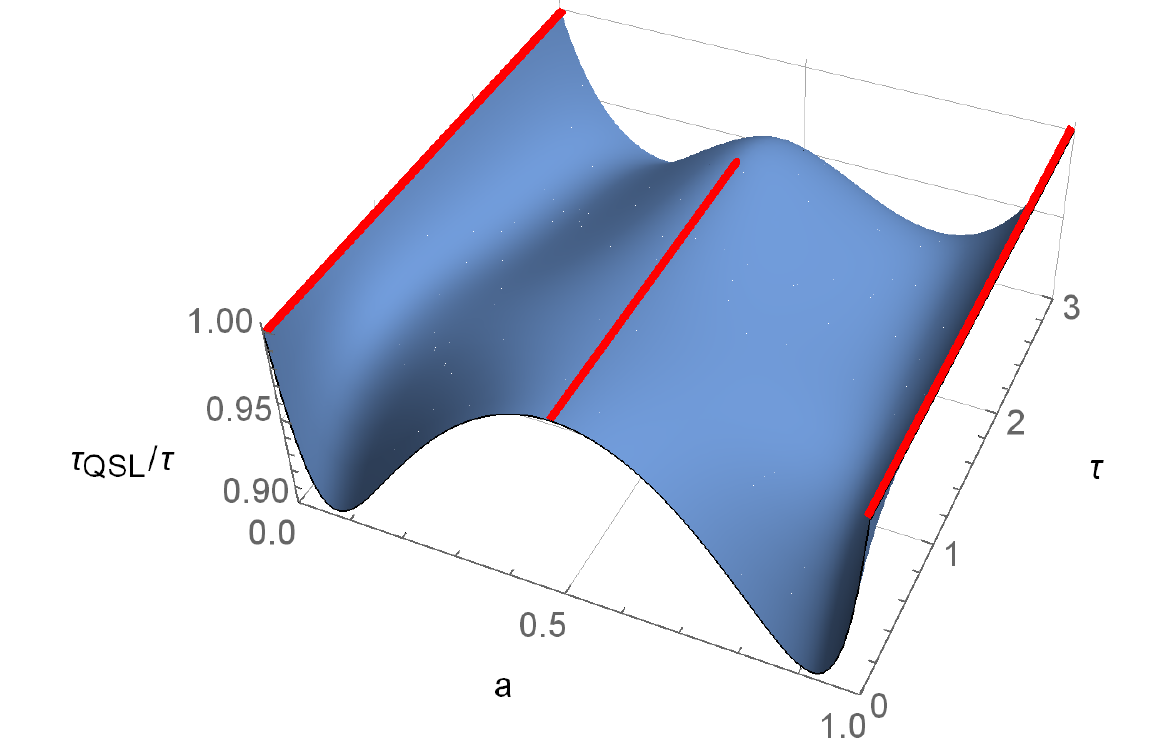}
\caption{The intial state and $\tau$ dependence of $\tau_{QSL}/\tau$ for the time-dependent system in Eq.~\eqref{tim-dep-model}. The optimal states are found at $a=0$ and $a=1$. Up to the point where $\tau = 2 \arctan (5/3)$, the choice $a=1/2$ results in $\tau_{QSL}/\tau = 1$, but drops down after it, see Fig.~\ref{fig:testME_t_vs_tQSL} for a detailed cross-section at $a=1/2$. The red highlights represent the points where the ratio is 1.}\label{fig:QSL_a_optimize_timedep}
\end{figure}

\begin{figure}[!h]\centering
\includegraphics[width=0.5\columnwidth]{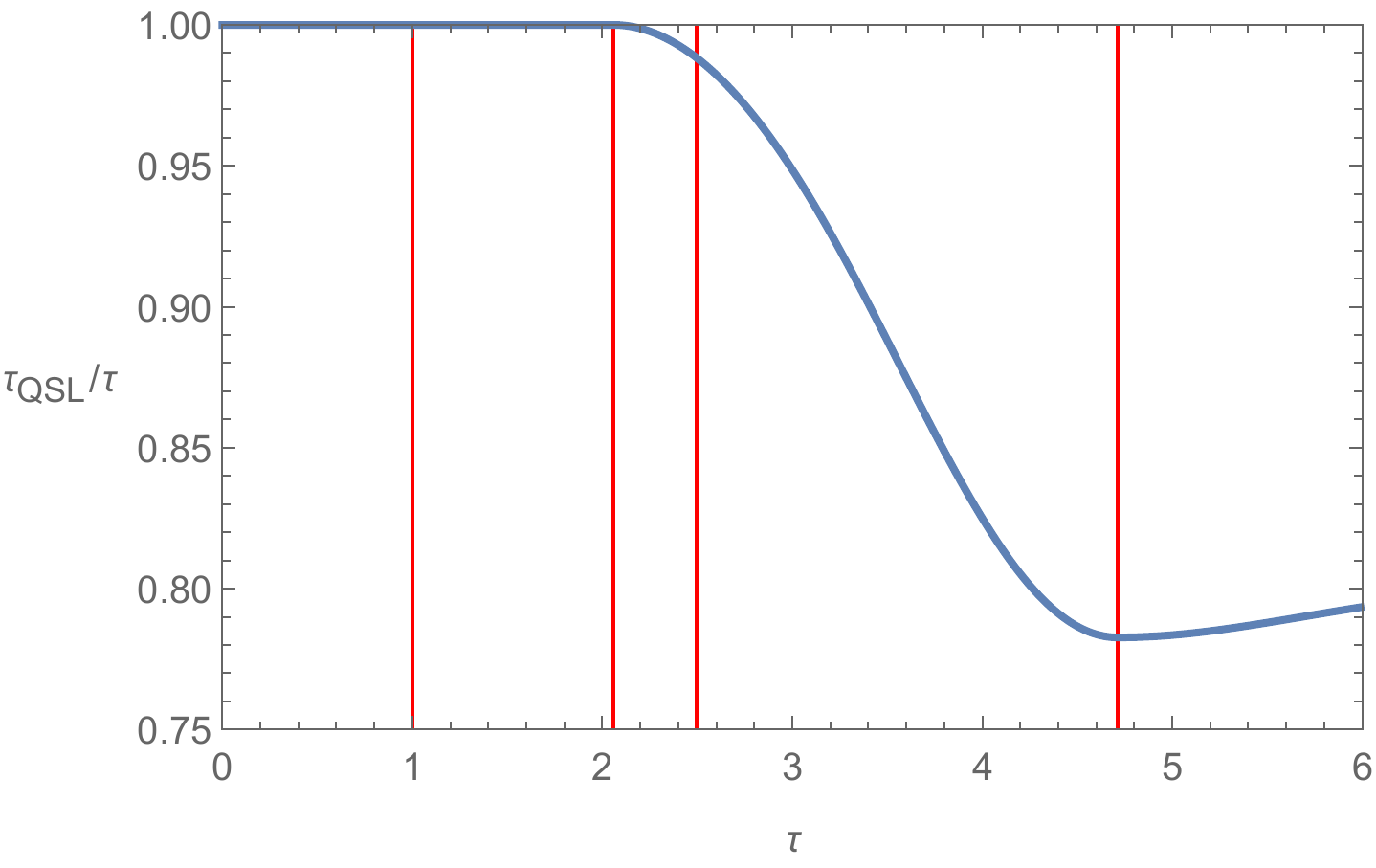}
\caption{The plot of $\tau_{QSL}/\tau$ as a function of $\tau$, with $a=1/2$. The red vertical lines represent the red points in Fig.~\ref{fig:Regionplot_with_testgamma} between $\tau = 0$ and $\tau = 6$. We see that $\tau_{QSL}/\tau = 1$ until $\tau = 2 \arctan (5/3)$, when the dynamics becomes BLP non-Markovian (see Fig.~\ref{fig:Regionplot_with_testgamma}). When the decay rates become positive again, that is at $\tau = 3\pi/2$, we see, that the QSL starts increasing again.}\label{fig:testME_t_vs_tQSL}
\end{figure}
\begin{figure}[!h]\centering
\begin{subfigure}[c]{\textwidth}
\centering\includegraphics[width=0.5\columnwidth]{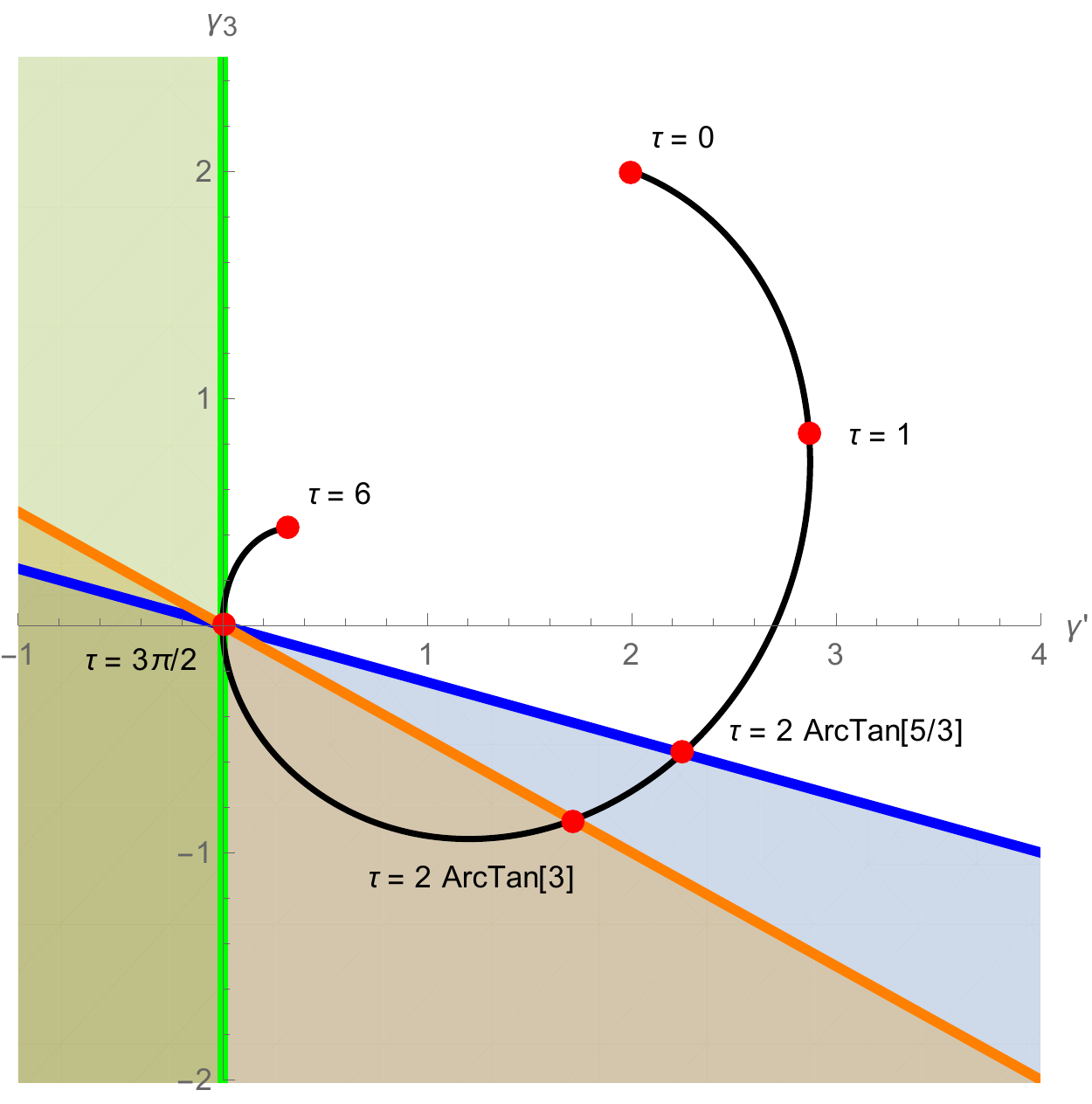}
\raisebox{25mm}{\includegraphics[width=0.25\columnwidth]{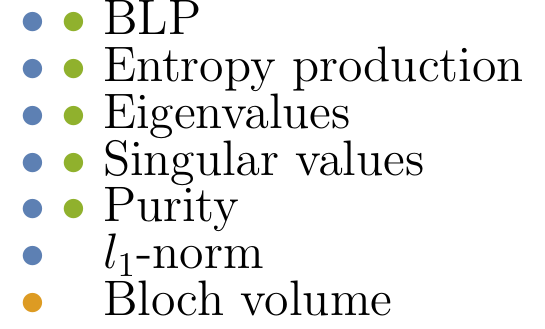}}
\end{subfigure}
\caption{Plot of the evolution of $\gamma'(t) \equiv \gamma_1(t) + \gamma_2(t)$ and $\gamma_3(t)$ in the $\lbrace \gamma',\gamma_3 \rbrace$ -space from $\tau = 0$ to $\tau = 6$. The times when the dynamics crosses a border are $2\arctan(5/3) \approx 2.061$, $2\arctan(3) \approx 2.498$, and $3\pi/2 \approx 4.712$. The lines represent the transition between Markovian and non-Markovian dynamics w.r.t.~different definitions of non-Markovianity. The lines representing the non-Markovianity conditions are: $\gamma'(t) - 4 \gamma_3(t) = 0$ (Blue), $\gamma'(t) - 2 \gamma_3(t) = 0$ (orange), and $\gamma'(t) = 0$ (green). The figure shows, that the measures of non-Markovianity connected to the blue line are critical for $\tau_{QSL}/\tau$ in this model, but the ones described by the orange line are not. The non-Markovianity indicators connected to the borders are listed on the right side of the figure. Colour indicates the non-Markovian region attached to that indicator. Multiple colors indicate that the non-Markovian region is represented by the union of these colours. For more details see Ref.~\cite{teittinen2018}.}
\label{fig:Regionplot_with_testgamma}
\end{figure}

In Fig.~\ref{fig:QSL_a_optimize_timedep} we see the initial state and evolution time dependence of $\tau_{QSL}/\tau$. We see that the initial state with $a=1/2$ is an optimal state up to $\tau = 2 \arctan (5/3)$. We note, that while the system is BLP non-Markovian, the optimal states $a=0$ and $a=1$ remain optimal. 

Fig.~\ref{fig:testME_t_vs_tQSL} shows how the change of $\tau$ affects $\tau_{QSL}/\tau$ for $a = 1/2$. We see that $\tau_{QSL}/\tau = 1$ until $\tau = 2\arctan (5/3)$. We also notice, that at $\tau = 3 \pi /2$, that is when $\gamma_1(t), \gamma_2(t) \gamma_3(t) \geq 0$ again, $\tau_{QSL}/\tau$ starts to increase. This is in accordance with Eq.~\eqref{eq:general_ratio_form}: When $\gamma_3(t)$ is positive, $\rho_\tau$ becomes less  similar with $\psi_0$ as $\tau$ increases. Thus, $F(\rho_\tau,\psi_0)$ decreases and as a consequence $\tau_{QSL}/\tau$ increases.

Fig.~\ref{fig:Regionplot_with_testgamma} shows the case of Fig.~\ref{fig:testME_t_vs_tQSL} in $\lbrace \gamma',\gamma_3\rbrace$ -space, where we have defined $\gamma'(t) = \gamma_1(t) + \gamma_2(t)$. The red lines of Fig.~\ref{fig:testME_t_vs_tQSL} are represented by the red dots in Fig.~\ref{fig:Regionplot_with_testgamma}. The coloured lines represent the border between Markovian and non-Markovian dynamics, as defined by different indicators of non-Markovianity, for the phase-covariant qubit master equation. The coloured region is where the dynamics is non-Markovian w.r.t.~the corresponding indicator. The union of blue and green regions is related to the BLP non-Markovianity, as well as non-Markovianity defined using entropy production, eigenvalues and singular values of the map and purity. The orange region is related to the Bloch volume indicator and does not concern our analysis of the BLP measure. For more details see Ref.~\cite{teittinen2018}. By comparing Figs.~\ref{fig:testME_t_vs_tQSL} and \ref{fig:Regionplot_with_testgamma}, we notice that the points where the $\tau_{QSL}/\tau$ changes dramatically in Fig.~\ref{fig:testME_t_vs_tQSL} coincides with the transition between Markovian and non-Markovian dynamics in Fig.~\ref{fig:Regionplot_with_testgamma}.

\section{Conclusions}\label{sec:conclusions}

In this paper, we have studied the connection between the quantum speed limit, the evolution time, non-Markovianity, and the initial state for a qubit system undergoing generic and several large sub-classes of dynamics.
We have derived general conditions for the optimal QSL bound, that is when $\tau_{QSL}/\tau = 1$, for a general qubit system, and studied some more special cases. Using these conditions, we studied the link between BLP non-Markovianity and QSL. We found, that in some cases, it is possible to generalize the results of \cite{xu2014} by showing that the QSL depends directly on the BLP measure. In general the connection becomes more complicated: We characterized classes of dynamics where the BLP Markovianity does not imply $\tau_{QSL}/\tau = 1$, and even cases where the QSL bound is tight for non-Markovian dynamics.

Despite concentrating on the BLP non-Markovianity, our analysis has implications to other definitions too:  Our results show that in some cases the tightness of the QSL bound is not achieved even for BLP Markovian dynamics (Fig.~\ref{fig:classes_of_dynamics} (A)), while in other cases BLP non-Markovianity is required for reaching the QSL bound (Fig.~\ref{fig:classes_of_dynamics} C (iii), C (iv), and D). As a consequence, there cannot exist a general connection between QSL and any definition of non-Markovianity which is in hierarchical relation with the BLP non-Markovianity either, as that would require such definition to be simultaneously both stronger and weaker than the BLP measure. (For a review of hierarchies between different definitions of non-Markovianity, see Ref.~\cite{li2018}.)

We have also shown that the QSL bound in an open qubit system is not tight for all pure initial states, even in purely Markovian systems. We analytically solved the optimal initial states leading to $\tau_{QSL}/\tau = 1~\forall~\tau\ge 0$ in dynamical semigroups rising from phase-covariant and Pauli master equations. We also studied the initial state dependence for example dynamics violating CP-divisibility. For all of the dynamical maps considered, the bound can be reached for a very few initial pure states, except for depolarizing dynamics.

Finally, we have analyzed the behaviour of the QSL across the Markovian to non-Markovian crossover, and found out that the tightness of the bound is clearly connected to the crossover in the example considered. In the non-Markovian region of the $\lbrace \gamma',\gamma_3 \rbrace$-space, the QSL starts to decrease. Conversely, when the dynamics becomes Markovian again, the QSL starts to increase but does not return to the optimal value $\tau_{QSL}/\tau = 1$. These results are in full accordance with our results concerning the connection between BLP non-Markovianity and QSL bound.

\section*{Acknowledgments}

The authors acknowledge financial support from the Academy of Finland Center of Excellence program (Project no. 312058) and the Academy of Finland (Project no. 287750). HL acknowledges also the financial support from the University of Turku Graduate School(UTUGS).



\section*{Bibliography}

\bibliographystyle{iopart-num}
\bibliography{NJP_110850_references}

\end{document}